# Observation of an Electric-Field Induced Band Gap in Bilayer Graphene by Infrared Spectroscopy

Kin Fai Mak[1], Chun Hung Lui[1], Jie Shan[2], and Tony F. Heinz[1*]

[1]*Departments of Physics and Electrical Engineering, Columbia University, 538 West 120$^{th}$ St., New York, NY 10027, USA*
[2] *Department of Physics, Case Western Reserve University, 10900 Euclid Avenue, Cleveland, OH 44106, USA*

Abstract

It has been predicted that application of a strong electric field perpendicular to the plane of bilayer graphene can induce a significant band gap. We have measured the optical conductivity of bilayer graphene with an efficient electrolyte top gate for a photon energy range of 0.2 – 0.7 eV. We see the emergence of new transitions as a band gap opens. A band gap approaching 200 meV is observed when an electric field ~ V/nm is applied by an electrolyte gate to the bilayer system, inducing a carrier density of about $10^{13}$ cm$^{-2}$. The magnitude of the band gap and the features observed in the infrared conductivity spectra are broadly compatible with calculations within a tight-binding model.

*Corresponding author: tony.heinz@columbia.edu

The system of bilayer graphene has emerged as an attractive material for fundamental studies of two-dimensional (2D) physics, as well as for many potential device applications [1-3]. In the bilayer system, the band structure arises from the coupling of two graphene monolayers, each of which would separately exhibit linearly dispersing conduction and valence bands that meet at the Dirac point [4, 5]. Coupling of the two monolayer graphene sheets in the usual A-B stacking of bilayer graphene yields pairs of hyperbolically dispersing 2D valence and conduction band that are split from one another by the interlayer interaction [6, 7]. This bilayer system shares many of the interesting properties of graphene, but provides a richer band structure. One important feature that is preserved is the absence of a band gap, *i.e.*, the upper valence band touches the lower conduction band at the K-point of the Brillouin zone. Recent theoretical studies have, however, predicted that a significant band gap could be induced by lowering the symmetry of the system through the application of a perpendicular electric field [1, 7]. One could thus produce a material with an electrically tunable band gap, a phenomenon of great significance for both basic physics and applications. The subject has accordingly generated much recent theoretical interest [8-15]. In particular, calculations indicate that band gaps of 100 meV or more can be induced in this manner [1, 7, 9]. Experiments involving chemical doping of graphene bilayers through the deposition of adlayers reveal the presence an appreciate band gap [1, 16]. To date, however, no direct evidence for the opening of a sizable band gap by electrostatic gating has been reported, although transport measurements have suggested a change in band structure and placed an upper bound of a few meV for an induced gap [2].

In this paper we report direct spectroscopic signatures of the opening of a large and tunable band gap in bilayer graphene induced by a perpendicular electric field. Using an electrolyte top gate to apply a strong electric field, we observe an induced band gap approaching 200 meV. This changed band structure is probed by infrared (IR) conductivity measurements. From the observation of optical transitions from states on both sides of the energy gap to the common higher-lying conduction (or lower-lying valence) band, the size of the induced band gap can be deduced directly from experiment. The current study complements earlier IR conductivity measurements of bilayer graphene under relatively weak electric fields that revealed changes arising from state-filling effects [17-19]. Our experimental results for the magnitude of the gap as a function of the applied electric field are consistent with theory based on a tight-binding (TB) picture of the electronic structure of graphene. This model also reproduces key features in the IR conductivity spectra, although the lack of precise quantitative agreement suggests that a more complete treatment of screening of the applied electric field [10, 15] and of many-body interactions [20-22] may be required. Our findings not only offer insight into the control of 2D electronic structure, but also immediately suggest the possibility of electronic and optoelectronic devices with greatly enhanced performance. In particular, a gap of the demonstrated magnitude is sufficient to produce room-temperature field-effect transistors with a high on-off ratio, something not possible in conventional graphene materials.

Large area (> 2000 $\mu m^2$) graphene bilayers were prepared on transparent $SiO_2$ substrates by mechanical exfoliation of kish graphite. They were identified based on their optical contrast [23, 24] and their thickness was confirmed by Raman spectroscopy [25]. Electrical contacts to the samples were made by electron-beam lithography and electron-beam evaporation of Au (50 nm). In addition to the standard source and drain contacts, an extra Au electrode was deposited within 100 μm of the bilayer sample to serve as a top gate through a transparent polymer matrix. The polymer electrolyte (poly(ethylene oxide):LiClO$_4$, 8:1, dissolved in methanol) [26] was cast onto the sample and dried at 110°C in ambient. The capacitance of such top gates (~ 1.0 F/cm$^2$) is nearly two orders of magnitude larger than that of standard Si/SiO$_2$ back gates [27] and, correspondingly, gives us access to much greater electric field strengths. The IR measurements on the bilayer devices were performed using a micro-Fourier Transform Infrared apparatus with a Globar source and a MgCdTe detector.

To extract the optical conductivity of the graphene bilayer, we first normalized the transmission spectrum of the bilayer by that of the reference taken in a region without the bilayer. For a thin film (the bilayer graphene) embedded between two transparent media (the polymer and the SiO$_2$ substrate), the film absorbance $A(\hbar\omega)$ for photon energy $\hbar\omega$ is given in terms of the normalized transmission $T(\hbar\omega)$ as $A(\hbar\omega) = [(n_{SiO_2} + n_{PEO})/2n_{PEO}][1 - T(\hbar\omega)]$, where $n_{PEO}$ = 1.45 [28] and $n_{SiO2}$ = 1.46 [29] are the refractive indices of the polymer (PEO) and the SiO$_2$ substrate, respectively [24]. We can express the results equivalently in terms of the real part of the optical sheet conductivity $\sigma(\hbar\omega)$ through the relation $\sigma(\hbar\omega) = (c/4\pi)A(\hbar\omega)$.

Measured conductivity spectra $\sigma(\hbar\omega)$ for a graphene bilayer under different top-gate bias voltages are displayed in Fig. 1. The corresponding source-drain current as a function of the top-gate voltage $V$ is shown in the inset of Figure 1(a). The charge neutrality point $V = V_{CN}$, identified by the minimum of the conductance, lies at - 0.5 V. Gate biases below and above this value induce hole [Fig. 1(a)] and electron [Fig. 1(b)] doping, respectively. The conductivity spectrum $\sigma(\hbar\omega)$ for $V = V_{CN}$ exhibits peaks at $\hbar\omega$ = 0.33 and 0.38 eV. As we increase $V$, we observe two significant changes in $\sigma(\hbar\omega)$ [Fig. 1(b)]. (i) The amplitude of the higher-energy transition (0.38 eV) grows, while that of the lower-energy transition (0.33 eV) subsides and disappears for $V-V_{CN} \approx 0.5$ V [Fig. 1(b)]. (ii) Absorption peaks, *P1* and *P2*, emerge from the higher-energy feature, becoming prominent for $V-V_{CN} > 0.5$ V. The two new peaks shift in opposite directions and broaden with increasing the gate bias. Corresponding effects are observed for hole doping [Fig. 1(a)].

To understand the observed optical properties of gated bilayer graphene, let us first consider its electronic structure. We present a TB description of the electronic structure of bilayer graphene under a perpendicular electric field [7], but the conclusions concerning the interpretation of the spectroscopic features do not depend on the details of the model. We include nearest intra- $\gamma_0$ and inter-layer $\gamma_1$ couplings and treat the applied electric field through the development of a self-consistent interlayer potential difference. As mentioned above, in the absence of an applied field (Fig. 2, red lines), the bilayer has pairs of split valence (*v1, v2*) and conduction bands (*c1, c2*) with energies $E^{(v1)}$, $E^{(v2)}$, $E^{(c1)}$, and $E^{(c2)}$. The pairs of valence and conduction bands are nearly parallel to one another and are separated by the energy $\gamma_1$. The system has no band gap, however, with bands *v1* and *c1* being degenerate at the K-point in the Brillouin zone. For low gate biases, the induced band gap is very slight and the electronic structure remains essentially unchanged. Electron doping (and analogously for hole doping) enhances transitions between bands *c1* and *c2,* as state filling provides more initial states for the optical transition. Correspondingly, transitions between bands *v1* and *v2* are suppressed by state blocking. Because of charge inhomogeneity in the sample from the electrolyte gate, both transitions (at slightly different energies due to electron-hole asymmetry [18, 19]) are seen simultaneously when $V \approx V_{CN}$. From the separation of the two resonances, we estimate the energy difference $\delta_{AB}$ = 25 meV between sub-lattices A and B within the same graphene layer, comparable to that reported by Li *et al.* [18].

The electronic structure of bilayer graphene change significantly when a strong electric field is applied, particularly in the region around the K point (Fig. 2, green curve). A gap develops between valence band *v1* and conduction band *c1*. The bands assume the slightly undulating shape of the so-called Mexican hat dispersion [7]. The conduction band minima and valence band maxima are shifted slightly away from the K- point, but for moderate values of the induced gap, the dispersion is weak and the band gap $E_g \approx \Delta E_K = E_K^{(c1)} - E_K^{(v1)}$, the energy gap at the K-point .

The origin of the peaks *P1* and *P2* observed experimentally is readily understood in terms of this modified band structure. They both arise (for electron doping) from transitions to the upper conduction band *c2*, with *P1* originating from the lower conduction band *c1* and *P2* from the upper valence band *v1*. The peaks reflect the high joint density of states present near the K-point; the progressive shift in energy with applied electric field comes from the growing gap in the bilayer electronic structure. This assignment of the features is supported by explicit calculations of $\sigma(\hbar\omega)$ presented below. The experimental energy difference $\Delta E$ between the two peaks *P2* and *P1* permits us to determine the induced band gap, since

$$\Delta E = E_{P2} - E_{P1} \approx [E_K^{(c2)} - E_K^{(v1)}] - [E_K^{(c2)} - E_K^{(c1)}] = E_K^{(c1)} - E_K^{(v1)} = \Delta E_K \approx E_g. \qquad (1)$$

The experimental results for $\Delta E \approx E_g$ as a function of gate voltages *V* (upper axis) are presented in Fig. 3. $\Delta E$ increases monotonically with increasing |V| and a value approaching 200 meV is observed for *V* = 4 V. As discussed below, these conditions correspond to a total induced charge $n \approx 1.5 \times 10^{13}$ cm$^{-2}$ and an effective (vacuum) electric-field strength above the graphene bilayer of 2.7 V/nm.

To compare our experimental results with theory, we need to convert the gate voltage *V* into the charge density *n* induced on the graphene bilayer. We do so by means of a relation that accounts both for changing of the gate capacitance $C_g$ and the shift in Fermi energy $\varepsilon_F$ of the bilayer sample [30]:

$$eV = \frac{ne^2}{C_g} + \varepsilon_F(n, \Delta). \qquad (2)$$

Here $\Delta$ denotes the interlayer potential difference parameter in the TB Hamiltonian [7], corresponding to the induced gap at the K-point, $\Delta E_K$. Using the TB expression for $\varepsilon_F(n, \Delta)$ [7] and choosing $C_g$ = 0.6 µF/cm$^2$ (corresponding to a Debye length of ~ 7 nm, typical for this type of electrolyte top gates), we then obtain *n* as a function of *V* (Fig. 3, lower horizontal axis compared with upper horizontal axis). This value of $C_g$ provides the best fit to the experimental IR conductivity spectra (discussed below) and is compatible with values obtained from independent measurements [31].

We can now examine the agreement between our experimental results for $\Delta E$ and the gap predicted from the self-consistent TB model developed by McCann [7]. Figure 3 presents a comparison of the variation of the experimental value $\Delta E$ with the predicted values of the gap at the K-point $\Delta E_K$ and the band gap $E_g$. The predicted band gap $E_g$ is also shown. In the calculation, we have used standard values for the TB parameters: interlayer coupling $\gamma_1 = 0.35$ eV, in-plane velocity $v = 1 \times 10^6$ m/s, interlayer distance $c_0 = 0.34$ nm, and dielectric screening within the bilayer $\varepsilon_r = 1$ [7]. We see that the TB model provides a reasonable description of the experimental data. For high biases there is a modest, but discernable difference between $\Delta E_K$ and $E_g$ as the fine structure of the Mexican hat dispersion becomes more prominent. This fact is reflected experimentally by broader absorption peaks of *P1* and *P2* and an expected intermediate location of $\Delta E$ between $\Delta E_K$ and $E_g$.

For a more detailed and direct comparison with the experimental data, we have simulated the IR conductivity spectra $\sigma(\hbar\omega)$ by means of the Kubo formula. We model the electronic structure within the TB model, using the dependence of the interlayer potential difference $\Delta$ on carrier density given by McCann [7]. We have also included a phenomenological broadening of 20 meV and an energy difference between sub-lattices A and B of $\delta_{AB} = 25$ meV. The simulation (Fig. 4) is seen to capture the main features of the experimental spectra (Fig. 1). With increasing bias, two separate peaks *P1* and *P2* emerge from a single feature and, as seen experimentally, shift in opposite directions. The peaks also broaden as a wider range of transitions in reciprocal space is allowed. In Fig. 4(c) we show for comparison the predicted $\sigma(\hbar\omega)$ under the neglect of any induced modification of the electronic structure or band gap opening. The behavior is completely inconsistent with experiment.

We see that a self-consistent application of simple TB theory yields reasonable overall agreement for the size of the gap (Fig. 3) and the principal features of the IR conductivity (Fig. 4 compared with Fig. 1) as a function of the bias voltage. However, modeling at the current level does not provide full quantitative agreement with experiment. While various factors may contribute to these discrepancies, two issues are of particular fundamental interest. One is the treatment of the screening of the applied electric field within the graphene bilayer. Recent *ab-initio* theory indicates the existence of more effective screening than included in TB models [10, 15]. This factor would influence the predicted magnitude of the gap (Fig. 3). Second, modeling of the IR conductivity at the TB level omits all many-body effects. This may account for discrepancy in the detailed spectral features of $\sigma(\hbar\omega)$, since many-body effects can play an important role in the optical properties of 2D electron gas systems [32] and carbon nanotubes [33, 34]. Comparison of the current experimental data with calculations incorporating many-body interactions should help to elucidate the nature of such excitonic effects in this model 2D system.

The authors acknowledge support from the National Science Foundation under grant CHE-0117752 at Columbia and grant DMR-0907477 at Case Western Reserve University; from DARPA under the CERA program; and from the New York State Office of Science, Technology, and Academic Research (NYSTAR).

**Figures:**

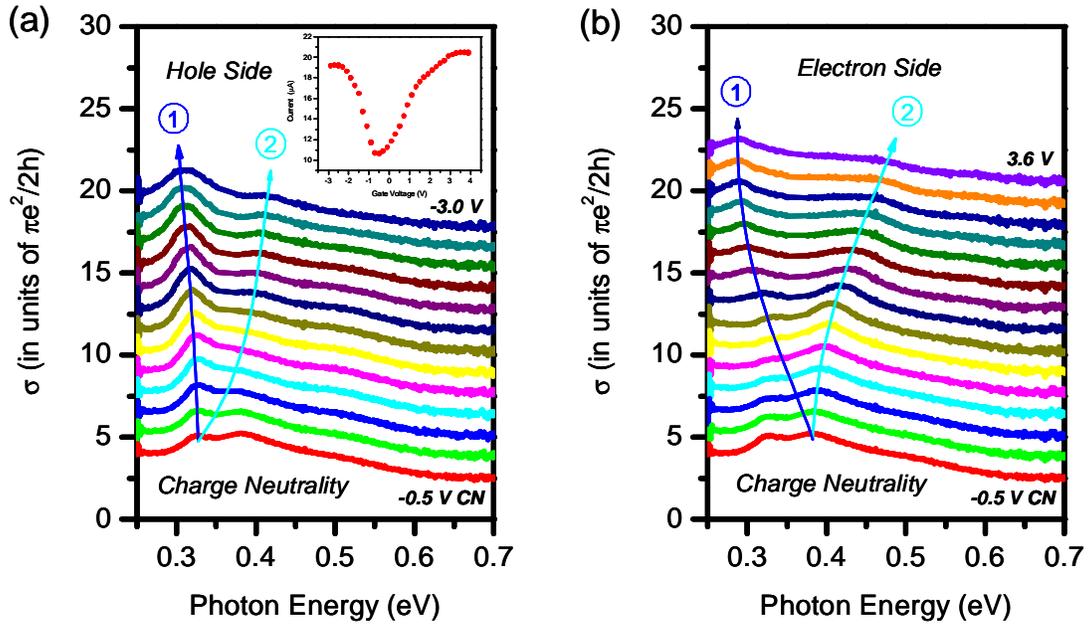

Fig. 1. Experimental spectra of the IR conductivity $\sigma(\hbar\omega)$ in units of $\pi e^2/2h$ (displaced vertically by 1.3 units from one another) of graphene bilayer under varying gate bias voltages $|V| < 4$ V. (a) corresponds to hole doping for $V$ = -0.5, -0.8, -1.0, -1.2, -1.4, -1.6, -1.8, -2.0, -2.2, -2.4, -2.6, -2.8 and -3.0 V (from bottom to top) and (b) to electron doping for $V$ = -0.5, -0.4, -0.3, -0.2, 0.0, 0.2, 0.4, 0.8, 1.2, 1.6, 2.0, 2.4, 2.8, 3.2 and 3.6 V (from bottom to top). Charge neutrality occurs for $V = V_{CN}$ = -0.5 V, as determined by the minimum in source-drain current [inset of (a)].

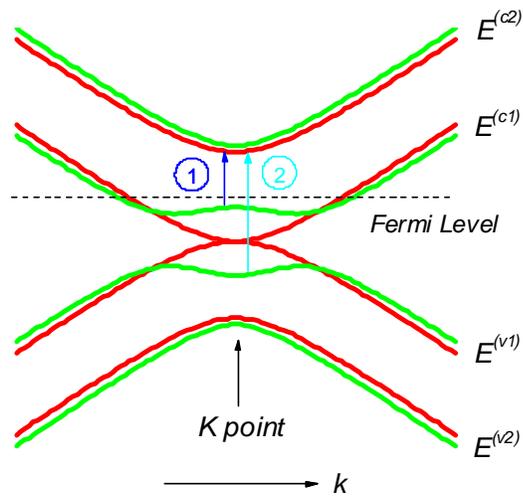

Fig. 2. Band structure of graphene bilayer with (green) and without (red) the presence of a perpendicular electric field as calculated within the TB model described in the text. Transitions 1 and 2 are the strongest optical transitions near the K-point for electron doping.

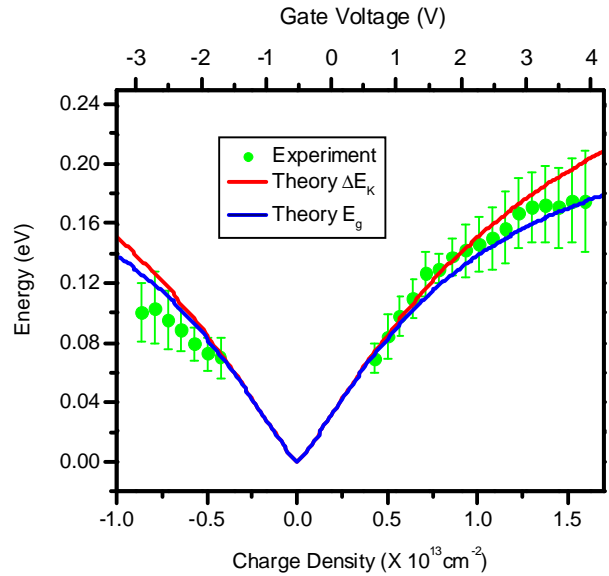

Fig. 3. Dependence of the energy gap at the K-point $\Delta E_K$ on the gate voltage and the charge doping density of the graphene bilayer. The error bars arise primarily from uncertainties in determining the peak position of the broad absorption features. The results of TB model for both $\Delta E_K$ and the band gap $E_g$ are plotted as well to compare with the experimental data.

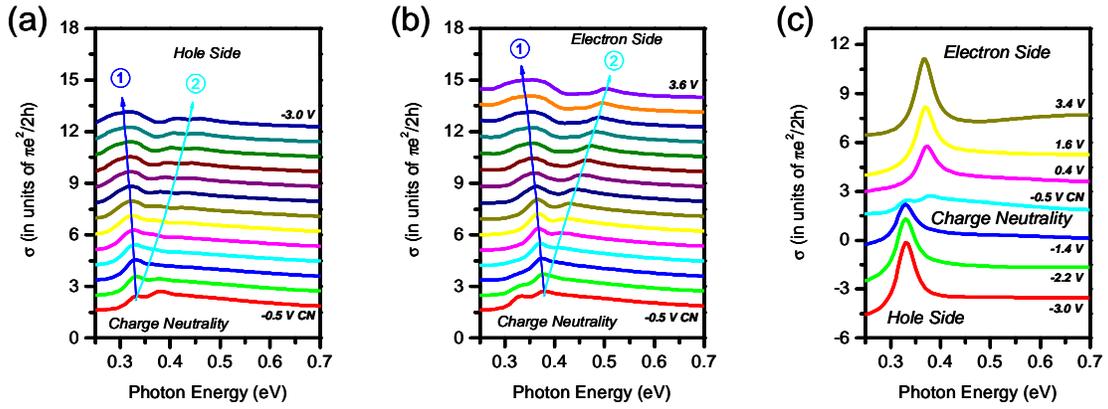

Fig. 4. Simulated spectra of the IR conductivity spectra $\sigma(\hbar\omega)$ spectra from the Kubo formula for gated bilayer graphene under the same doping as in Fig. 1. (a) and (b) show the results obtained with the predictions of TB model for the electronic structure described in the text (including the electron-hole asymmetry). For comparison, (c) is a reference calculation in which the bilayer graphene band structure is assumed to remain unchanged (without an induced gap) and the spectral modifications reflect only population changes.